\documentstyle[prl,aps,epsf,here]{revtex} 
\twocolumn 

\def\bgrk#1{\mbox{{\boldmath $#1$ \unboldmath}}\!\!}

\begin{document}
\draft
\title{Reply to a Comment by J. Bolte, R. Glaser and S. Keppeler on:\\
Semiclassical theory of spin-orbit interactions using spin coherent
       states}
\author{M. Pletyukhov, Ch. Amann, M. Mehta$^2$ and M. Brack}
\address{Institut f\"ur Theoretische Physik, Universit\"at Regensburg,
         D-93040 Regensburg, Germany\\
         $^2$present address: Harish-Chandra Research Institute, 
         Chhatnag Road, Jhusi, Allahabad 211019, India}
\maketitle

\noindent
The essential points in a recent Comment by Bolte, Glaser and Keppeler 
(BGK) \cite{bgk} are their claims ($i$) that our results in \cite{plet} 
contradict earlier 
findings by Bolte and Keppeler (BK) \cite{boke}, ($ii$) that a semiclassical 
trace formula with explicit coupling of orbital and spin degrees of freedom 
can only be obtained in the limit of infinite spin: $S\rightarrow\infty$, 
and ($iii$) that our approach \cite{plet} uses an incorrect 
application of the stationary-phase approximation. We disagree with all
three points.

$Ad~i\!:)$  We have already stated in \cite{plet} that in the ``weak-coupling 
limit'' (WCL), our approach yields the same result as that obtained in 
\cite{boke}.
The detailed proof is given in \cite{oleg}. In the WCL, the orbital motion is 
not affected by the spin, as found also in \cite{boke}, but it is essential 
in our approach to keep the $\hbar$-dependent terms in the principal function 
${\cal R}$, see Eq.\ (7) of \cite{plet}, and to take into account the 
Solari-Kochetov phase correction \cite{koch,sola}. Of the two versions of 
semiclassical trace formulae mentioned by BGK, the WCL thus leads to perfect 
agreement between our approach and that of BK \cite{boke}, and there is no 
contradiction. The ``strong-coupling limit'' (SCL) \cite{lifl,frgu} has not 
been studied from within our approach, so that no contradiction can be claimed.

$Ad~ii\!\!:)$ We have not ``overlooked" (as suggested by BGK) that there is a
formal problem with the use of a finite spin $S$, but clearly pointed this 
out in the last paragraph of our paper \cite{plet}. We have stated there that 
the path integral has the correct measure only in the large-spin limit, and 
that this calls for a proper renormalization scheme for finite spin. For pure
spin systems, the scheme is known \cite{koch,sola} and leads \cite{ston} to 
the so-called ``Weyl shift'' $S \rightarrow S + 1/2$ that yields a valid 
semiclassical description also in our formalism. It can, e.g., easily be 
checked that for a pure spin Zeeman interaction $\,-\mu\,\hat{\bgrk{\sigma}}
\cdot {\bf B}\,$ 
this semiclassical treatment becomes exact. (This also contradicts the
third objection by BGK, see point $iii$.) In spin-orbit coupled 
systems, the renormalization scheme is not known yet. However, as shown in 
\cite{oleg} and mentioned in point $i$, the Weyl shift appears to be justified 
at least in the weak-coupling limit, since it 
yields the same result as \cite{boke}. Therefore, there is good reason to 
expect our approach to give reasonable approximations for finite $S$ also in 
the general case -- as substantiated by the successful application to a real 
physical system in \cite{plet}.\\ A general remark is appropriate here. 
Even if a semiclassical approach is mathematically only justified in the limit 
of large quantum numbers like $S \rightarrow\infty$, one is entitled as a 
physicist to try and use it in situations where this limit is not 
fulfilled. It is a well-known bonus of semiclassical approximations that 
they work even in limits where they mathematically ought not to work 
(provided that the dependence on the quantum numbers is sufficiently smooth 
and that appropriate phase corrections -- the Maslov indices -- are 
incorporated). See, e.g., the WKB quantization of orbital motion 
which rigorously is justified only for quantum numbers $n\rightarrow \infty$, 
but in harmonic oscillators becomes exact even for $n = 0$.

$Ad~iii\!:)$ We do not see why our use of the stationary-phase approximation
should be incorrect. 
Whenever ${\cal R}_0\gg\hbar$, so that the stationary-phase method 
is justified, no harm is done in adding a small term of order $\hbar$ to 
${\cal R}_0$: the phase $({\cal R}/\hbar)$ for ${\cal R} = {\cal R}_0 + 
\hbar\,{\cal R}_1$ will still be rapidly oscillating. In fact, it is 
precisely the $\hbar$ terms in ${\cal R}$ (7) -- if properly treated, as 
shown in \cite{oleg} -- that yield the BK spin modulation factor in the 
weak-coupling limit (cf.\ point $i$). Another proof of the validity of
the stationary-phase approximation using (7) is the fact already mentioned
in point $ii$ that it leads to exact semiclassical results for pure spin
systems. {\it It is the very essence of our 
approach to have gone beyond the leading-order $\hbar$ approximation used 
by BK \cite{boke}}. To include $\hbar$ terms -- which are all proportional 
to $S$ -- into ${\cal R}$ (7) and the equations of motion (9), and hence 
to get a non-trivial spin-orbit coupling where the orbital motion is 
modified by the spin motion (which is physically sound), appears to be a 
well-working semiclassical approach. We do not claim it to be the only one.

Let us finally point out that for the system investigated in \cite{plet}
-- a quantum dot with Rashba spin-orbit interaction -- neither the WCL 
nor the SCL can be used: in the WCL the BK approach \cite{boke} leads to 
a trivial spin modulation factor 2 ignoring spin-orbit coupling effects, and 
the SCL is obstructed by the mode conversion problem \cite{lifl,frgu,cham}. 
This demonstrates that for certain systems a more general approach is 
required; we have proposed one that works in the above system. Another case 
where both WCL and SCL give wrong results is the two-dimensional 
electron gas with Rashba interaction in a homogeneous magnetic field. 
It was shown analytically in \cite{cham} that the BK trace formula for 
this system is correct only to leading order in the spin-orbit coupling
strength $\kappa$.

We are grateful to Oleg Zaitsev and Stephen Creagh for helpful discussions.

\end{document}